\documentclass[epj,final]{svjour}
\usepackage[dvipsnames]{xcolor}
\usepackage{graphicx}
\usepackage{float}
\usepackage{textcomp}
\usepackage{latexsym}
\usepackage{url}
\usepackage{amsfonts}
\usepackage{amsmath, amssymb}
\RequirePackage{graphicx}
\usepackage{caption,subcaption}
\usepackage{indentfirst}

\begin{document}
	\title{Neural Network Analysis of S2-Star Dynamics: Extended mass}
        \titlerunning{Neural Network Analysis of S2-Star Dynamics}
	\author{N.Galikyan\inst{1,2}, Sh.Khlghatyan\inst{2}, A.A.Kocharyan\inst{3}, V.G.Gurzadyan\inst{2,4,5}
	}                     
	%
	%
	\institute{National Research Nuclear University MEPhI, Moscow, Russia \and Center for Cosmology and Astrophysics, Alikhanian National Laboratory and Yerevan State University, Yerevan, Armenia  \and School of Physics and Astronomy, Monash University, Clayton, Australia \and SIA, Sapienza Universita di Roma, Rome, Italy \and e-mail: gurzadyan@yerphi.am (corresponding author)}
	\date{Received: date / Revised version: date}
	%

	\abstract{Physics-informed neural network (PINN) analysis of the dynamics of S-stars in the vicinity of the supermassive black hole in the Galactic center is performed within General Relativity treatment. The aim is to reveal the role of possible extended mass (dark matter) configuration in the dynamics of the S-stars, in addition to the dominating central black hole's mass. The PINN training fails to detect the extended mass perturbation in the observational data for S2 star within the existing data accuracy, and the precession constraint indicates no signature of extended mass up to $0.01\%$ of the central mass inside the apocenter of S2. Neural networks analysis thus confirm its efficiency in the analysis of the S-star dynamics.   
}

	\PACS{
		{98.80.-k}{Cosmology} 
	} 
%
\maketitle

\section{Introduction}

The study of the dynamics of S-stars in the Galactic center provides a unique opportunity to test General Relativity (GR) 
\cite{Grav,Genz} and to reveal the role of various physical processes in the very vicinity of a supermassive black hole. The S-star data have been used to constrain modified gravity models and of associated structures \cite{Cap,Amor,Fermionic,Borka,Z1,Z2,Yukawa,EinMax,GRAVITYscalarclouds,Lau}, thus acting as one more window with respect to other tests of GR and gravity theories, e.g. \cite{Ciu1,Ciu2,Ciu3,Event}. Among other applications of the S-star data is the search of the contribution  of extended dark matter configurations to the star dynamics \cite{Rub,Nuc,Z3,Abut,Z4,Chan,Hei,Lec}.

In our previous study \cite{PINNSstars}, we analyzed the S-stars motion using the so-called physics-informed neural networks (PINN) \cite{PINN1,PINN2} to test the weak-field modified General Relativity \cite{G} with the metric tensor
\begin{equation} 
g_{00} = 1 - \frac{2 G M}{r c^2} - \frac{\Lambda r^2}{3};\,\,\, g_{rr} = \left(1 - \frac{2 G M}{r c^2} - \frac{\Lambda r^2}{3}\right)^{-1},
\end{equation} 
of the Schwarzschild - de Sitter form \cite{R}. As weak-field modified GR, it was shown to describe the dynamics of the groups and clusters of galaxies, and the Hubble tension as a manifestation of two flows, local and global ones \cite{GS1,G1,GS5,SKG1,GS4}.  
In \cite{PINNSstars}, using the PINN we had determined the orbital parameters of the S1, S2, S9, S13, S31, S54  stars, and found constraint on the $\Lambda$-gravity by means of the dynamics of S2 star, which was monitored up to completing a single orbit \cite{Chan}.
Among other effects occurring in the vicinity of massive black holes surrounded by dense stellar systems is the  tidal disruption of stars \cite{GO1,GO2,Rees}.
   
Below, we apply PINN to reveal the extended mass signature in the data of S2 star motion data. We obtain an upper limit for the presence of an extended mass as one part in $10^{4}$ of the mass of central black hole within S2 apocenter.

\section{The method}

To study the correspondence between the motion data of the S2 star and the extended mass, we use PINN. Our model consists of two parts:
\begin{itemize}
    \item A neural network with dense layers in which the input are the polar angles $\varphi$ and the output is the inverse radius $u:=\tfrac{1}{r}$. The output is compared with the observed data and minimizes the mean squared error between them.
    \item The physical part includes differential equations corresponding to our physical model and provides an additional contribution to the loss function. If the physical process is given by the following equation
\begin{equation} \label{eq:diff_eq}
    F(x, y, y'_x, y_x'', \ldots, y_x^{(n)}) = 0,
\end{equation}
then the physical loss is defined by the following loss function
\begin{equation} \label{eq:phys_loss}
    L_{phys}(f(x), x) = F^2\left(x, f(x), f'(x), f''(x), \ldots, f^{(n)}(x)\right).
\end{equation}
\end{itemize}

We are considering the motion of S2 star using the individual training scheme introduced in \cite{PINNSstars}.

To proceed on the performance of the models we use the following metrics 
\begin{equation}\label{eq:Metric}
\begin{split}
    &\mathcal{M}_{\text{model-data}}=\mathbb{E}\left[1-\frac{|u_{\text{Model}}-u_{\text{Star}}|}{u_{\text{Star}}}\right],\\
    &\mathcal{M}_{\text{data-physics}} = \mathbb{E}\left[1 - \frac{|u_{\text{Phys}}-u_{\text{Star}}|}{u_{\text{Star}}}\right],\\
    &\mathcal{M}_{\text{model-physics}} = \mathbb{E}\left[1 - \frac{|u_{\text{Model}}-u_{\text{Phys}}|}{\frac{1}{2}(u_{\text{Model}}+u_{\text{Phys}})}\right],\\
\end{split}
\end{equation}
where $u_{\text{Model}}$ is the prediction of the model, $u_{\text{Star}}$ is the used data, and $u_{\text{Phys}} = \frac{1}{\hat{p}}(1+\hat{e}\cos(\varphi-\varphi_0))$ is the prediction by the obtained orbital parameters.

\subsection{The physical model}

In addition to the central point mass, we consider an additional component, namely, an extended mass configuration determined by the following density distribution
\begin{equation}\label{eq:Rho_em} 
    \rho(r) = \frac{\alpha}{4\pi} \left(\frac{r}{r_0}\right)^{-\gamma}, 
\end{equation}
so that the center is located at the foci of the elliptical orbit \cite{Z3}, i.e. coinciding with the central point mass, $r_0$ is a scaling radius of the extended mass (dark matter). The power-law form for the density run for the extended mass enables to cover various models, see \cite{Abut,Hei}. Parameter $\alpha$ of dimension $[\text{au}]^{-2}$ (we use geometrized unit system $G=c=1$) at $\alpha = 0 [\text{au}]^{-2}$ implies that there is no extended mass, and the radius $r_0$ is normalized so that $\alpha = 1 \text{au}^{-2}$ corresponds to an extended mass equal to the point mass $M_{\bullet}$.  The total mass of the system has the following form
\begin{equation}\label{eq:M_tot}
    M_{\text{tot}} = M_{\bullet} + M_{\text{EM}}(r),
\end{equation}
where $M_{\bullet}$ is the central point mass and the extended mass is
\begin{equation}\label{eq:M_extended}
    M_{\text{EM}}(r)=\left[\begin{aligned}&\alpha r_0^3 \ln\frac{r}{r_{\text{g}}},\quad \gamma = 3,  \\&\frac{\alpha}{3-\gamma }r_0^{\gamma}(r^{3-\gamma}- r_{\text{g}}^{3-\gamma}),\quad \gamma \neq 3,\end{aligned}\right.
\end{equation}
and $r_{\text{g}}:=2M_{\bullet}$ is the Schwarzschild radius, $\alpha$  determines the significance of the influence of an extended mass.

We are considering the motion in the Schwarzschild metric assuming that the value of extended mass is less than of the central point mass, as follows from the constraint obtained in \cite{Abut} i.e. the extended mass within the S2 star apocenter has to be less than 0.1\% of the mass of the central black hole. 

\begin{figure}[H]
    \centering
    \includegraphics[width=0.8\linewidth]{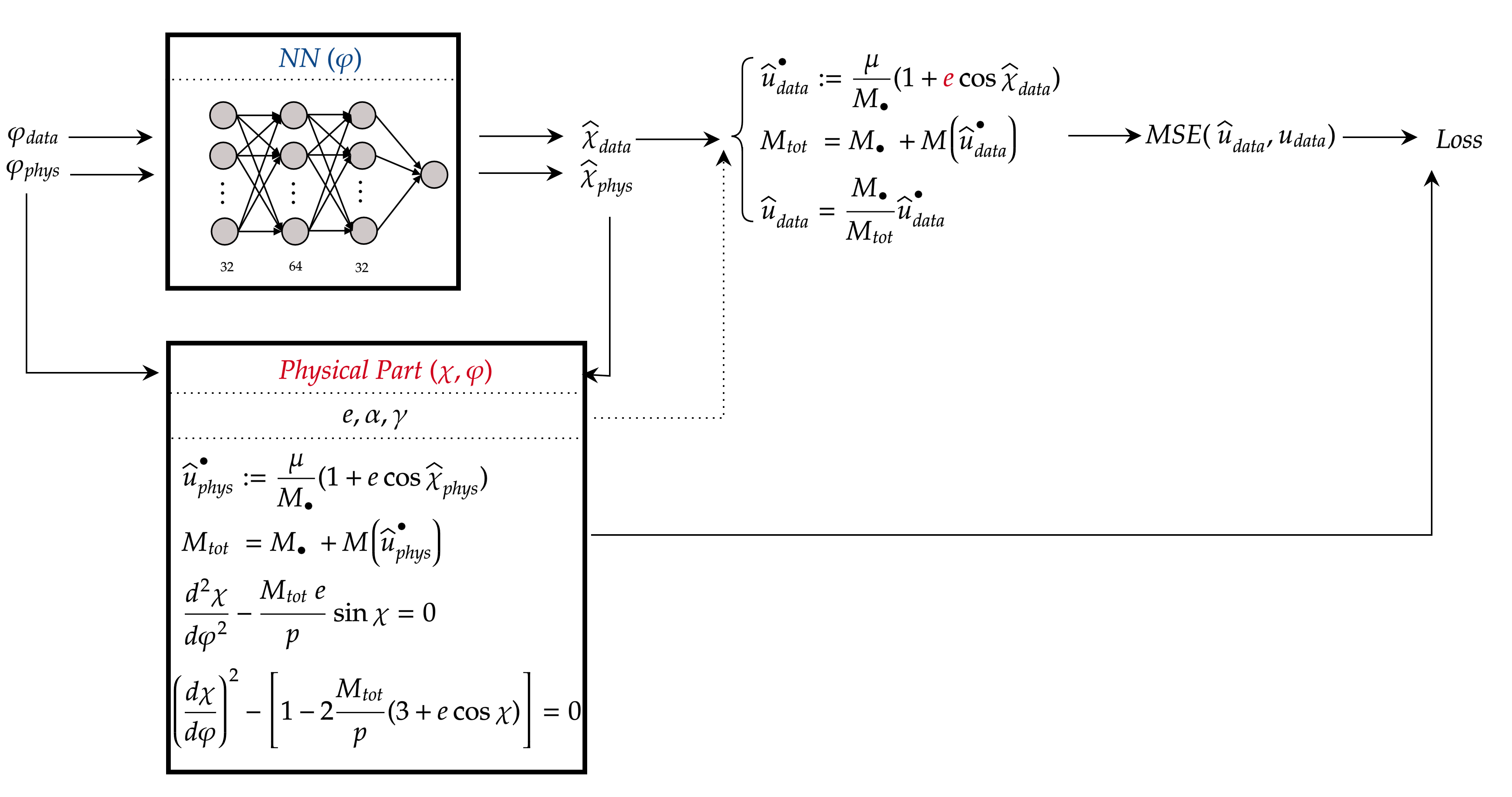}
    \caption{Model scheme}
    \label{fig:scheme}
\end{figure}

Using the Darwin's variable $\chi$ the equation of motion becomes \cite{Chandra}
\begin{align}\label{eq:Darwin}
\frac{d^2\chi}{d\varphi^2}&=\mu e \sin\chi,\\
\left(\frac{d\chi}{d\varphi}\right)^2&=1-2\mu(3+e\cos\chi),\\
u =\frac{\mu}{M}&(1+e\cos\chi),\,\, \mu:=\frac{M}{p},
\end{align}
where $u:=\tfrac{1}{r}$. Since we assume that the extended mass is much smaller than the point mass, we can approach the functional dependence of $u$ on $M_{\text{EM}}(u)$ in the following way. First, determine $u_{\bullet}$ by taking into account only the point mass $M_{\bullet}$ and after that compute the variable $u$, i.e.
\begin{align}
    u_{\bullet}=\frac{\mu}{M_{\bullet}}(1+e\cos\chi),\\
    u=(1-M_{\text{EM}}/M_{\bullet})u_{\bullet}.
\end{align}
We can also use the exact equation of motion by initially adding extended mass in the Lagrangian, but of course, the equation will not be obtained in the form of Darwin variable.
The network scheme is presented in Fig.(\ref{fig:scheme}).

\subsection{PINN training}

Our equations of motion Eq.(\ref{eq:Darwin}) contain five parameters $(M_{\bullet},e,p,\alpha,\gamma)$ or the $(M_{\bullet},e,\mu,\alpha,\gamma)$. We fix the parameters $p,M_{\bullet}$ obtained from the previous train \cite{PINNSstars} and leave only $e,\alpha, \gamma$ as  trainable parameters. 

We use an individual training scheme and gradually reduce the learning rate during the training. In addition, we select normalization coefficients for data and physical losses, so that the network is not focused only on solving a regression problem or on differential equations.

After training for trainable parameters, the following results were obtained:
\begin{align}
    &\hat{e}= 0.8859,\quad 
    \hat{\alpha}=-2.07\times 10^{-7},\quad 
    \hat{\gamma}= 0.2482, \\
    &\mathcal{M}_{\text{model-data}} = 0.9857 ,\quad \mathcal{M}_{\text{data-physics}} = 0.9860,\quad \mathcal{M}_{\text{model-physics}} = 0.9897.
\end{align}

\begin{figure}[H]
	\centering
	\begin{subfigure}{0.49\textwidth}
		\includegraphics[width=\linewidth,height=5.2cm]{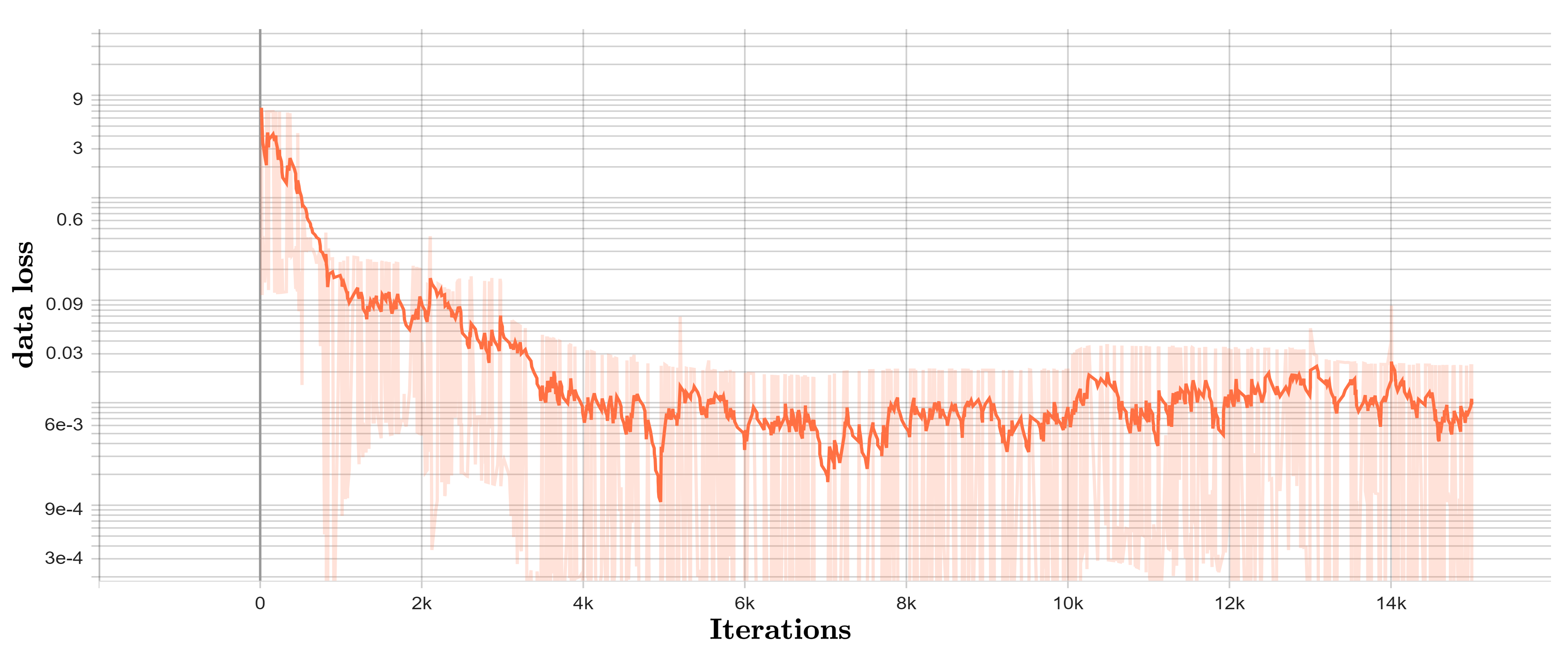}
		\caption{}
	\end{subfigure}
    \begin{subfigure}{0.49\textwidth}
		\includegraphics[width=\linewidth,height=5.2cm]{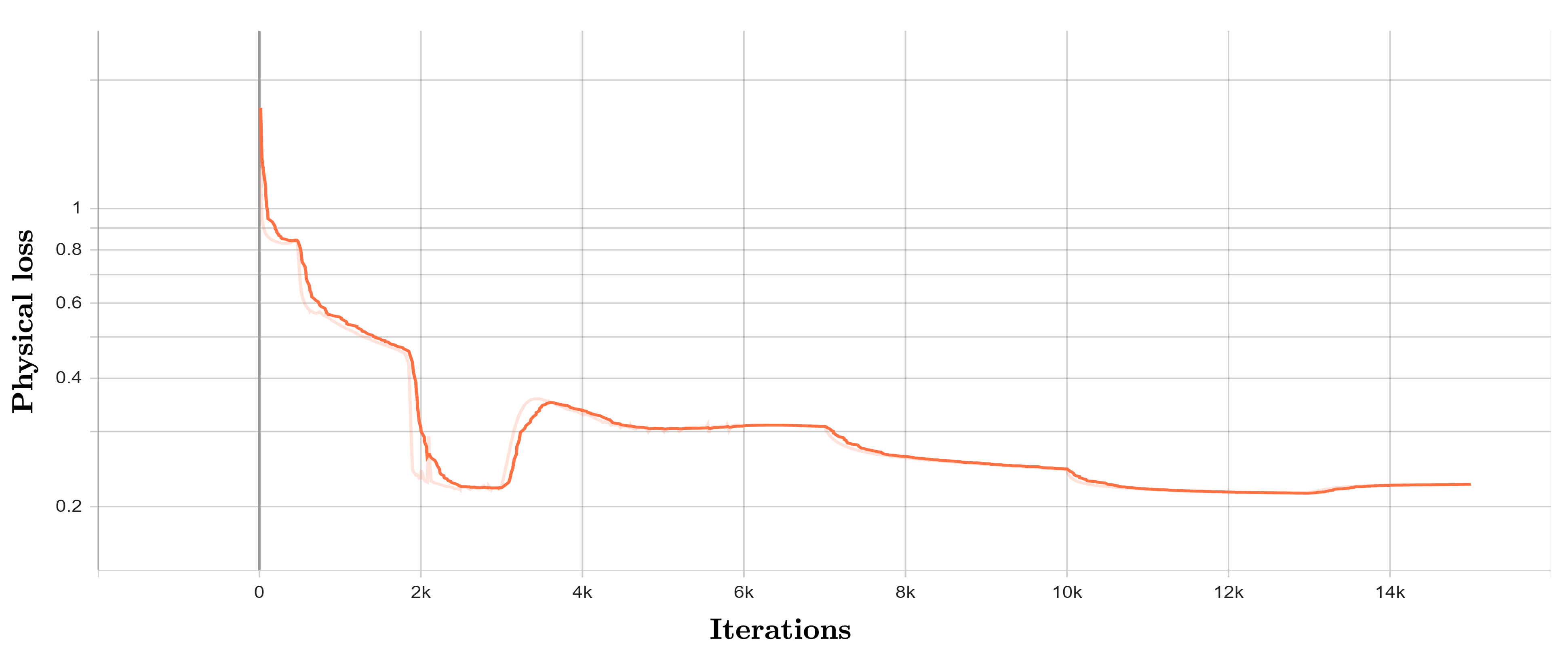}
		\caption{}
	\end{subfigure}\\
 \begin{subfigure}{0.49\textwidth}
		\includegraphics[width=\linewidth,height=5.2cm]{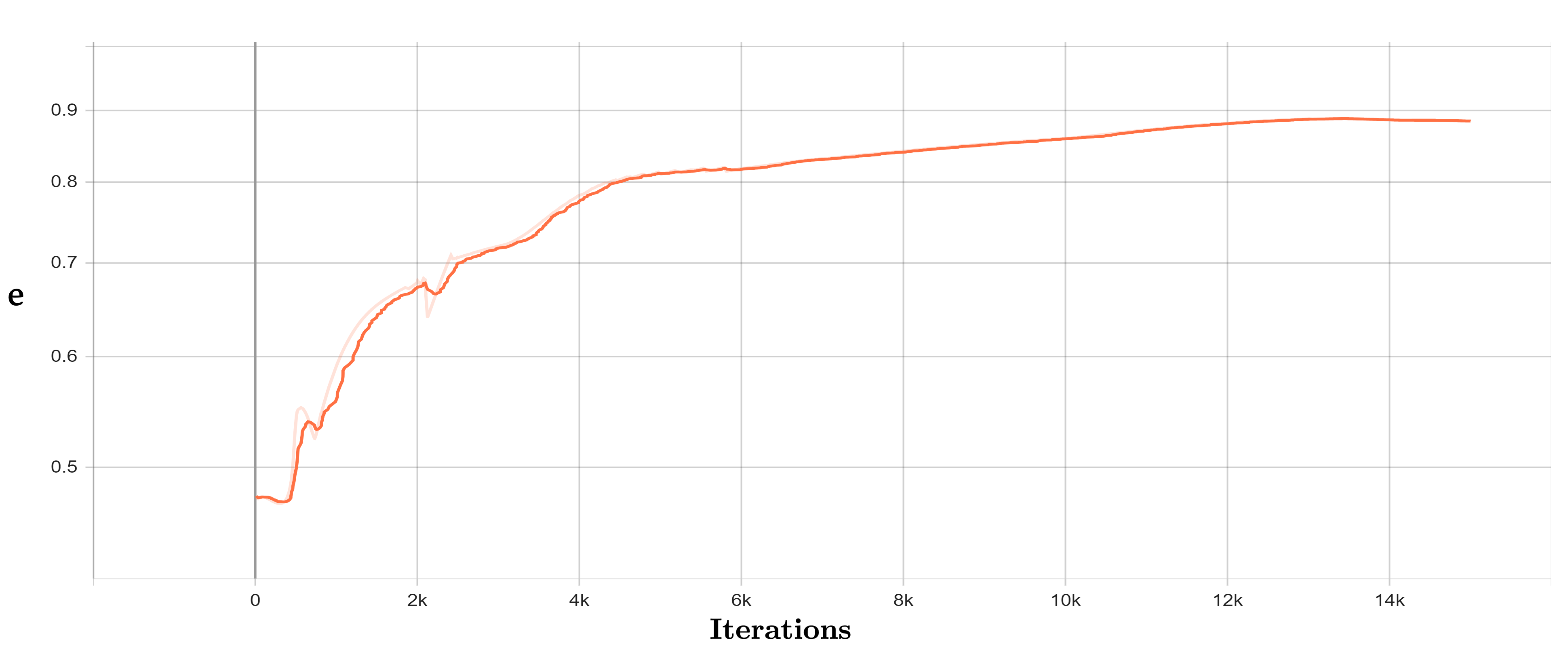}
		\caption{}
	\end{subfigure}
    \begin{subfigure}{0.49\textwidth}
		\includegraphics[width=\linewidth,height=5.2cm]{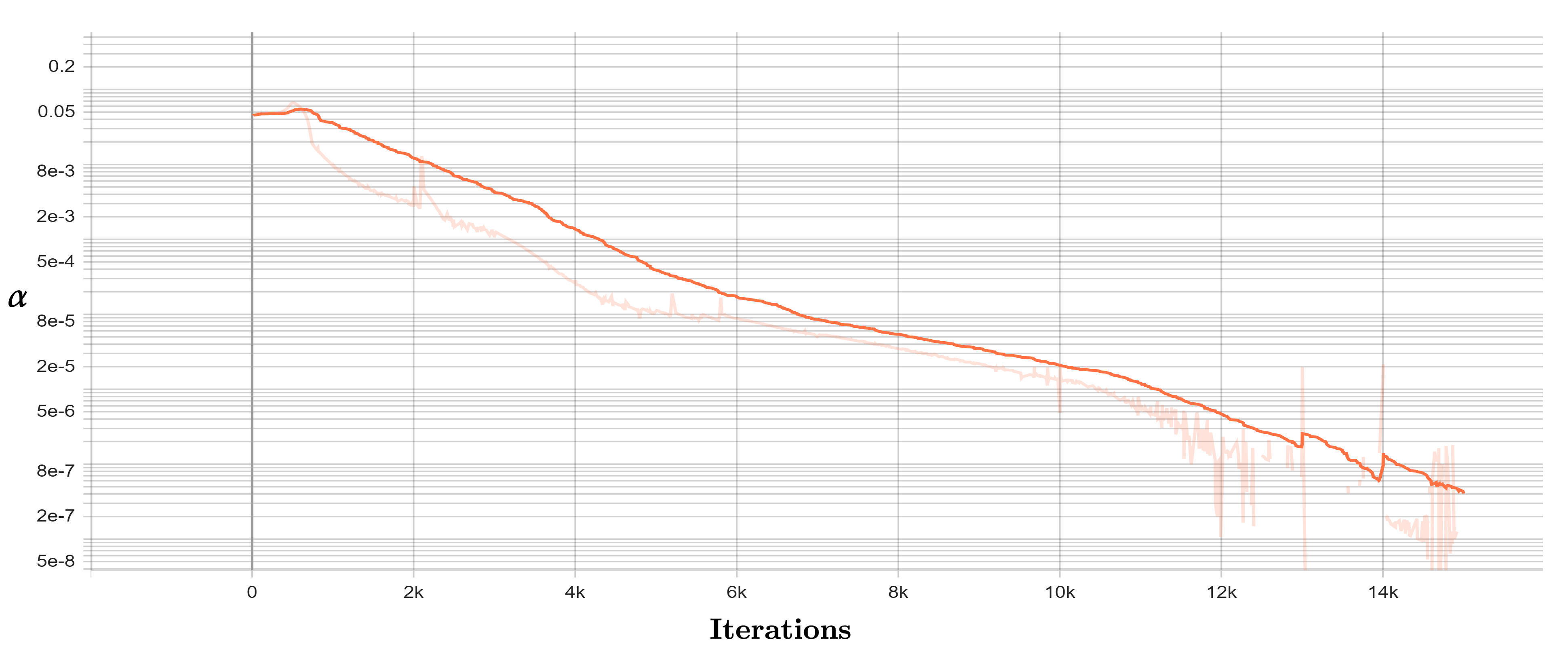}
		\caption{}
	\end{subfigure}
 \caption{The loss functions and the physical parameters $e, \alpha$ during training.}
 \label{fig:plots}
\end{figure}

\begin{figure}[H]
    \centering
    \includegraphics[width=1\linewidth]{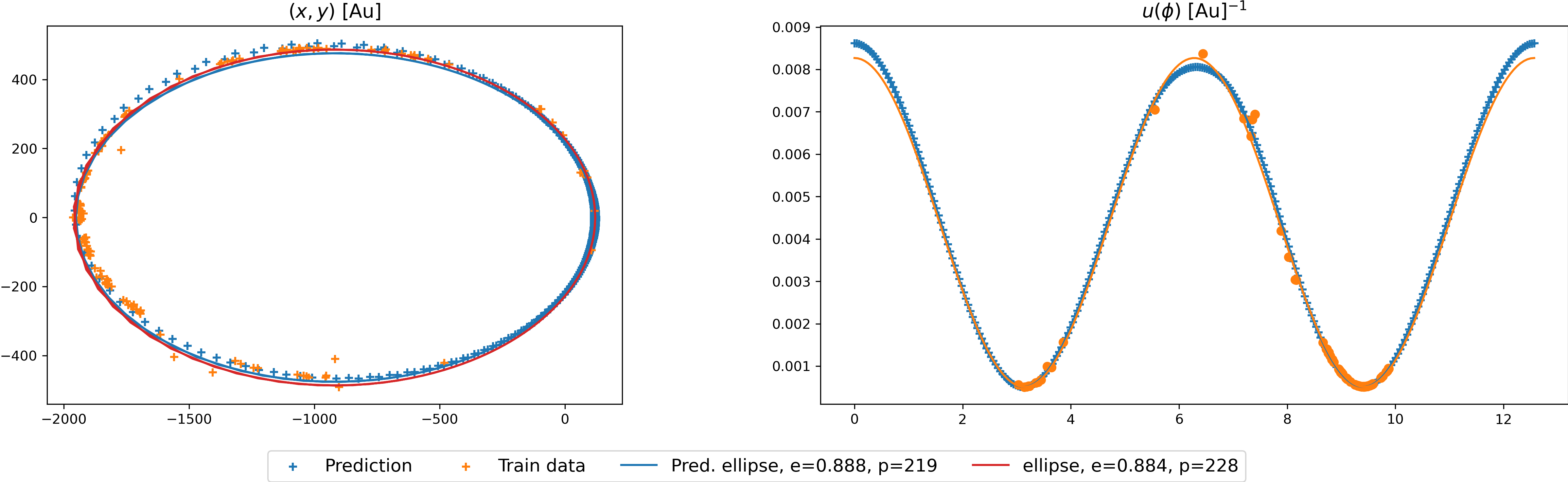}
    \caption{\textit{Left}: The predicted ellipse with $\hat{e}=0.888,\hat{p}=219\,\,[\text{au}]$ (blue line) and  ellipse with $e=0.884,\hat{p}=228\,\,[\text{au}]$ \cite{Gillessen} (orange line). \textit{Right}: Regression result.}
    \label{fig:results}
\end{figure}

Changes in the loss function and physical parameters during training are shown in Fig.\ref{fig:plots} and the predicted trajectory and regression result are shown in the Fig.\ref{fig:results}. 
From the Fig.(\ref{fig:plots}) we see that the value of $\alpha$ is practically zero and much less than the loss value, thus, the network does not "see" the presence of the $\alpha$ parameter. The value of $\alpha$  may further decrease, but this will in no way affect the loss values. In addition the reached value of the $\gamma$ parameter is not the specific value. This can be verified by initializing the $\gamma$ to a different value or during training artificially changing its value.

The constraint on $\alpha$ can be found based on precession, similar to the case of the modified gravity constraint \cite{PINNSstars}.

\section{Precession \& constraint on $\alpha$}

As in the case of $\Lambda$-gravity \cite{PINNSstars}, also in this case it is possible to find a constraint on $\alpha$ attributing it as responsible for an additional precession. The extended mass leads to an additional precession, which can be found if we consider the perturbation to the Keplerian potential
\begin{equation}
    V(r)=-\frac{M_{\bullet}}{r}+\delta V_{\text{GR}}(r)+\delta V_{\text{EM}}(r), 
\end{equation}
where $\delta V_{\text{GR}}(r) = \frac{r_{\text{g}}L^2}{2}\frac{1}{r^3}$, is $\delta V_{\text{EM}}(r)$ is to be found from the equation
\begin{equation*}
    -\frac{M_{\text{EM}}(r)}{r^3}\boldsymbol{r} = -\nabla V_{\text{EM}}(r),
\end{equation*}
and leading to the following differential equation
\begin{equation}
    \frac{d}{dr}\delta V_{\text{EM}}(r) = \frac{M_{\text{EM}}(r)}{r^2}.
\end{equation}
Then the precession rate is given by
\begin{equation}
    \delta\varphi_{\text{EM}}=\frac{\partial}{\partial L}\left(\frac{1}{L}\int_{0}^{2\pi}r^2\delta V_{\text{EM}}(r)d\varphi\right),
\end{equation}
where $L$ is the angular momentum and the first integral for unperturbed case.

For the $\gamma=1$ and $\gamma=4$ we get the following expressions, respectively
\begin{align}
    \label{eq:alpha_prec1}
    \delta\varphi_{\text{EM}} &= -\frac{\pi}{2\sqrt{2}}\frac{L}{|E|^{3/2}}\alpha r_0\qquad \gamma=1,\\ 
    \label{eq:alpha_prec4}
    \delta\varphi_{\text{EM}} &=-\frac{\pi}{L^2}\alpha r_0^4 \qquad \gamma = 4,
\end{align}
where $E$ is the orbital energy. To calculate the constraint, $E$ and $L$ must be expressed in terms of orbital parameters.  Using the results of the previous training: $\delta\varphi_{\text{Reg}}$ and $\delta\varphi_{\text{Phys}}$
\begin{align}
    &\delta\varphi_{\text{Reg}}= 11.84';\quad\sigma_{\text{Reg}}=0.03',\\
    &\delta\varphi_{\text{Phys}}=11.82';\quad\sigma_{\text{Phys}}=0.02',
\end{align}
we find the constraint on $\alpha$, by calculating the minimal precession (both (\ref{eq:alpha_prec1}) and (\ref{eq:alpha_prec4}) are negative) as $(\delta\varphi_{\text{Phys}} + 3\sigma_{\text{Phys}}) - (\delta\varphi_{\text{Reg}} - 3\sigma_{\text{Reg}})$:
\begin{align}
    \alpha &\leq 3.9\cdot10^{-14} { \text{[au]}^{-2}},  \qquad \gamma = 1, \\
    \alpha &\leq 3.8\cdot10^{-4} { \text{[au]}^{-2}},  \qquad \gamma = 4.
\end{align}
The fraction of the extended mass at the apocenter  $r_{\text{a}}:=\frac{\hat{p}}{1-\hat{e}}$ then yields
\begin{align}
   \frac{M_{\text{EM}}(r_{\text{a}})}{M_{\bullet}} &\leq 7.7\cdot10^{-7},  \qquad \gamma = 1, \\
    \frac{M_{\text{EM}}(r_{\text{a}})}{M_{\bullet}}&\leq 5.5\cdot10^{-4},  \qquad \gamma = 4.
\end{align}

\section{Conclusions}

We used the neural network PINN to study if the S2-star data leave a room for extended mass distribution in the vicinity of the central black hole. Considering that the extended mass is less than the central point mass as indicated by earlier studies \cite{Abut} and assuming the density has a power-law form, and also taking into account that the center of the extended mass should be located in the center of the galaxy, we modify the relativistic equation of motion (in Darwin notations \cite{Chandra}) accordingly.

PINN training shows that the network does not "see" the parameter $\alpha$  in the observed data, thus indicating that the found $\alpha$-mass itself is much less than the total loss.

Then, based on the fact that the extended mass has to contribute to the S2 precession, we found a constraint on the $\alpha$ parameter using the result of previous training without extended mass. The calculations show that the contribution of the extended mass depends on the density power-law index $\gamma$, but is again too small as compared to the point mass.

Thus, the neural network analysis of the S2-star data aimed to reveal the possible contribution of the extended mass component within the existing data accuracy does not indicate any signature of extended mass at least up to one part of $10^{4}$  of the central mass within S2 apocenter.

\section{Acknowledgments}
We are thankful to the referee and G. Hei\ss el for useful comments. Sh.K. is acknowledging the ANSEF grant 23AN:PS-astroth-2922.

\section{Data Availability Statement} 
Data sharing not applicable to this article as no datasets were generated or analysed during the current study.

\newpage
\end{document}